\documentclass{elsart}

\usepackage[square,comma]{natbib}
\usepackage{graphicx}
\usepackage{pxfonts}
\usepackage{lineno}

\usepackage{amssymb}

\journal{}

\begin{document}
\thispagestyle{empty}
\begin{Large}
\textbf{DEUTSCHES ELEKTRONEN-SYNCHROTRON}

\textbf{\large{Ein Forschungszentrum der
Helmholtz-Gemeinschaft}\\}
\end{Large}

DESY 10-008

January 2010

\begin{eqnarray}
\nonumber &&\cr \nonumber && \cr \nonumber &&\cr
\end{eqnarray}
\begin{eqnarray}
\nonumber
\end{eqnarray}
\begin{center}
\begin{Large}
\textbf{Ultrafast X-ray pulse measurement method}
\end{Large}
\begin{eqnarray}
\nonumber &&\cr \nonumber && \cr
\end{eqnarray}

\begin{large}
Gianluca Geloni,
\end{large}
\textsl{\\European XFEL GmbH, Hamburg}
\begin{large}

Vitali Kocharyan and Evgeni Saldin
\end{large}
\textsl{\\Deutsches Elektronen-Synchrotron DESY, Hamburg}
\begin{eqnarray}
\nonumber
\end{eqnarray}
\begin{eqnarray}
\nonumber
\end{eqnarray}
ISSN 0418-9833
\begin{eqnarray}
\nonumber
\end{eqnarray}
\begin{large}
\textbf{NOTKESTRASSE 85 - 22607 HAMBURG}
\end{large}
\end{center}
\clearpage
\newpage

\begin{frontmatter}



\title{Ultrafast X-ray pulse measurement method}


\author[XFEL]{Gianluca Geloni\thanksref{corr},}
\thanks[corr]{Corresponding Author. Tel: ++49 40 8998 5450. Fax: ++49 40 8998 1905. E-mail address: gianluca.geloni@xfel.eu}
\author[DESY]{Vitali Kocharyan,}
\author[DESY]{and Evgeni Saldin}

\address[XFEL]{European XFEL GmbH, Hamburg, Germany}
\address[DESY]{Deutsches Elektronen-Synchrotron (DESY), Hamburg,
Germany}

\begin{abstract}
In this paper we describe a measurement technique capable of
resolving femtosecond  X-ray pulses from XFEL facilities. Since
these ultrashort pulses are themselves the shortest event
available, our measurement strategy is to let the X-ray pulse
sample itself. Our method relies on the application of a "fresh"
bunch technique, which allows for the production of a seeded X-ray
pulse with a variable delay between seed and electron bunch. The
shot-to-shot averaged energy per pulse is recorded. It turns out
that one actually measures the autocorrelation function of the
X-ray pulse, which is related in a simple way to the actual pulse
width. For implementation of the proposed technique, it is
sufficient to substitute a single undulator segment with a short
magnetic chicane. The focusing system of the undulator remains
untouched, and the installation does not perturb the baseline mode
of operation. We present a feasibility study and we make
exemplifications with typical parameters of an X-ray FEL.
\end{abstract}

%
%

\end{frontmatter}



\section{\label{sec:intro} Introduction and method}

\begin{figure}[tb]
\includegraphics[width=1.0\textwidth]{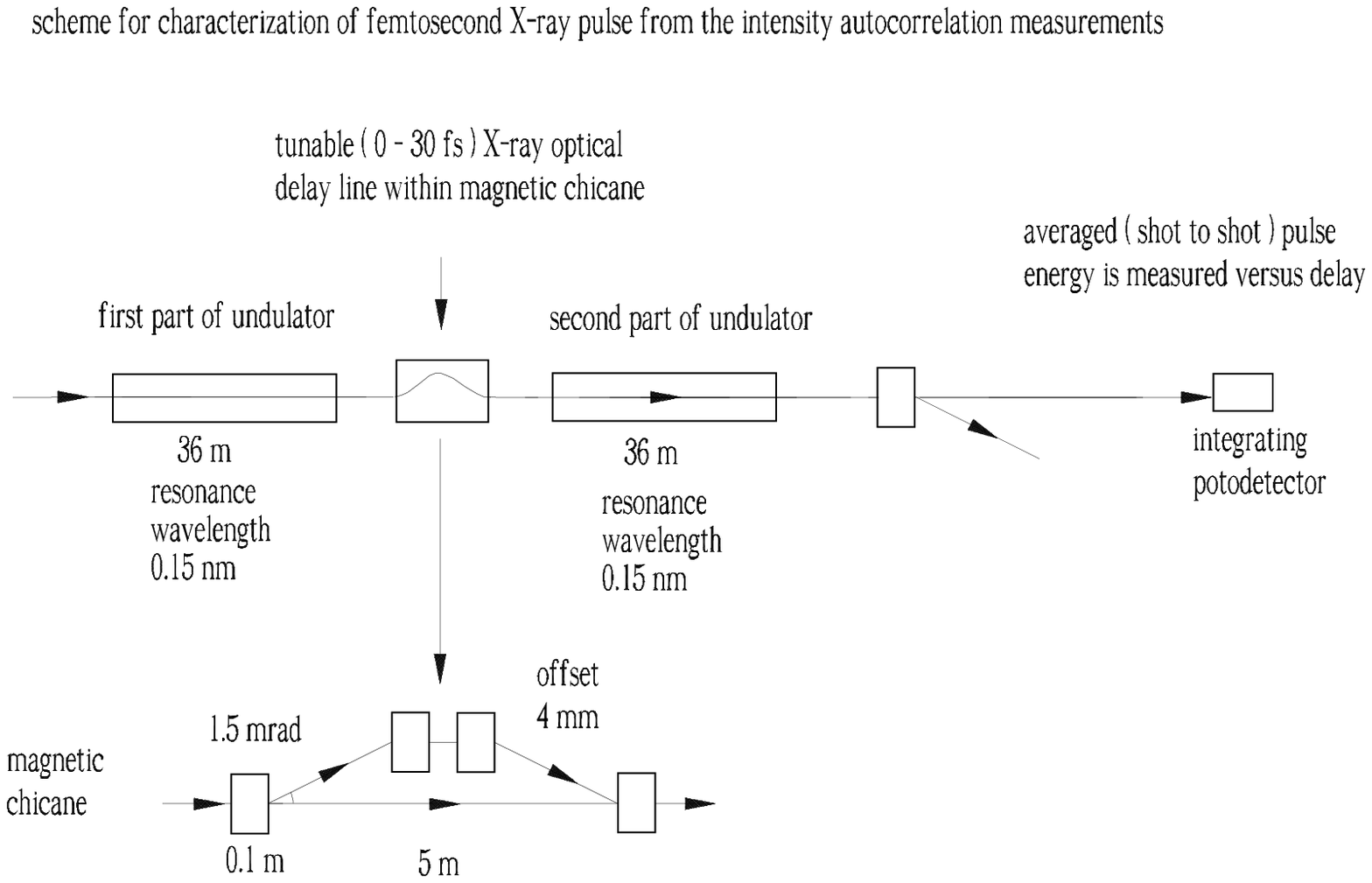}
\caption{Experimental layout for ultra-short  X-ray
pulse-measurement using "fresh" bunch technique.  Pulse and
electron bunch are separated at the exit of first part of
undulator, one is variably delayed with respect to the other.
Modulation of the electron bunch is washed out. X-ray pulse and
"fresh" electron bunch are overlapped in the second part of
undulator. The averaged X-ray pulse energy at the exit of  setup
is measured versus delay, yielding the autocorrelation trace. }
\label{mm1}
\end{figure}
\begin{figure}[tb]
\begin{center}
\includegraphics[width=100mm]{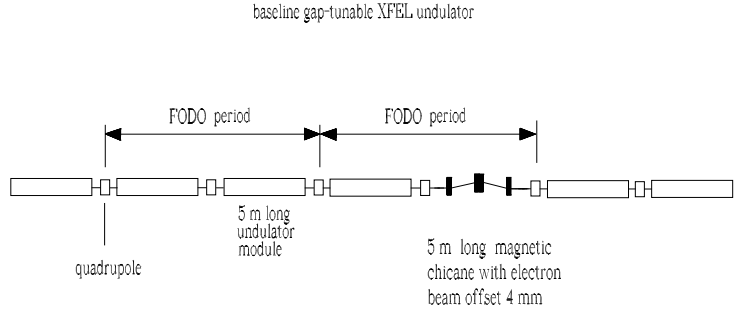}
\caption{\label{delaymag} Installation of a magnetic delay in the
baseline XFEL undulator.}
\end{center}
\end{figure}
\begin{figure}[tb]
\includegraphics[width=1.0\textwidth]{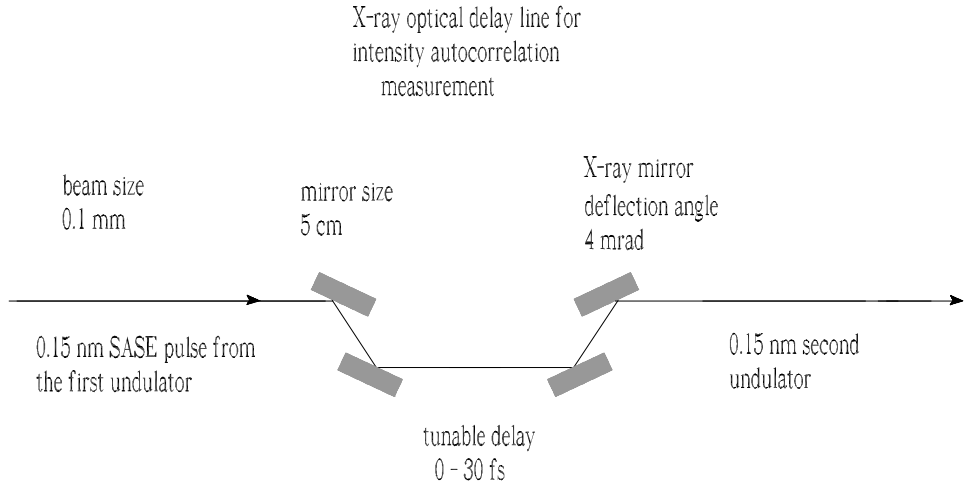}
\caption{X-ray optical system for delaying the SASE pulse with
respect to the electron bunch. The X-ray optical system can be
installed within the magnetic chicane between first and second
part of the undulator.} \label{mm4}
\end{figure}

The measurement of X-ray pulses on the femtosecond time scale
constitutes an unresolved problem. It is possible to create
sub-ten femtosecond X-ray pulses from XFELs, but not to measure
them. In fact, conventional photodetectors and streak-camera
detectors do not have a fast enough response time to characterize
ultrashort radiation pulses. For example, the rise time of the
best streak-cameras approaches $100$ fs, far too slow to resolve
femtosecond pulses. Special measurement techniques are needed. In
this paper we propose a new method for the measurement of the
duration of femtosecond X-ray pulses from XFELs. The method is
based on the measurement of the autocorrelation function of the
X-ray pulses. The setup in Fig. \ref{mm1}, similar to that
described in \cite{OUR01},  may be used to this purpose. The
electron bunch enters the first part of the baseline undulator and
produces SASE radiation with ten MW-level power. After the first
part of the undulator, the electron bunch is guided through a
short magnetic chicane whose function is both, to wash out the
electron bunch modulation, and to create the necessary offset to
install an X-ray optical delay line. The chicane is short enough
to be installed in the space of a single XFEL segment, as shown in
Fig. \ref{delaymag}, and does not perturb the focusing structure
of the machine. The optical delay line is sketched in Fig.
\ref{mm4}, and has already been discussed in \cite{OUR02}.

\begin{figure}[tb]
\includegraphics[width=1.0\textwidth]{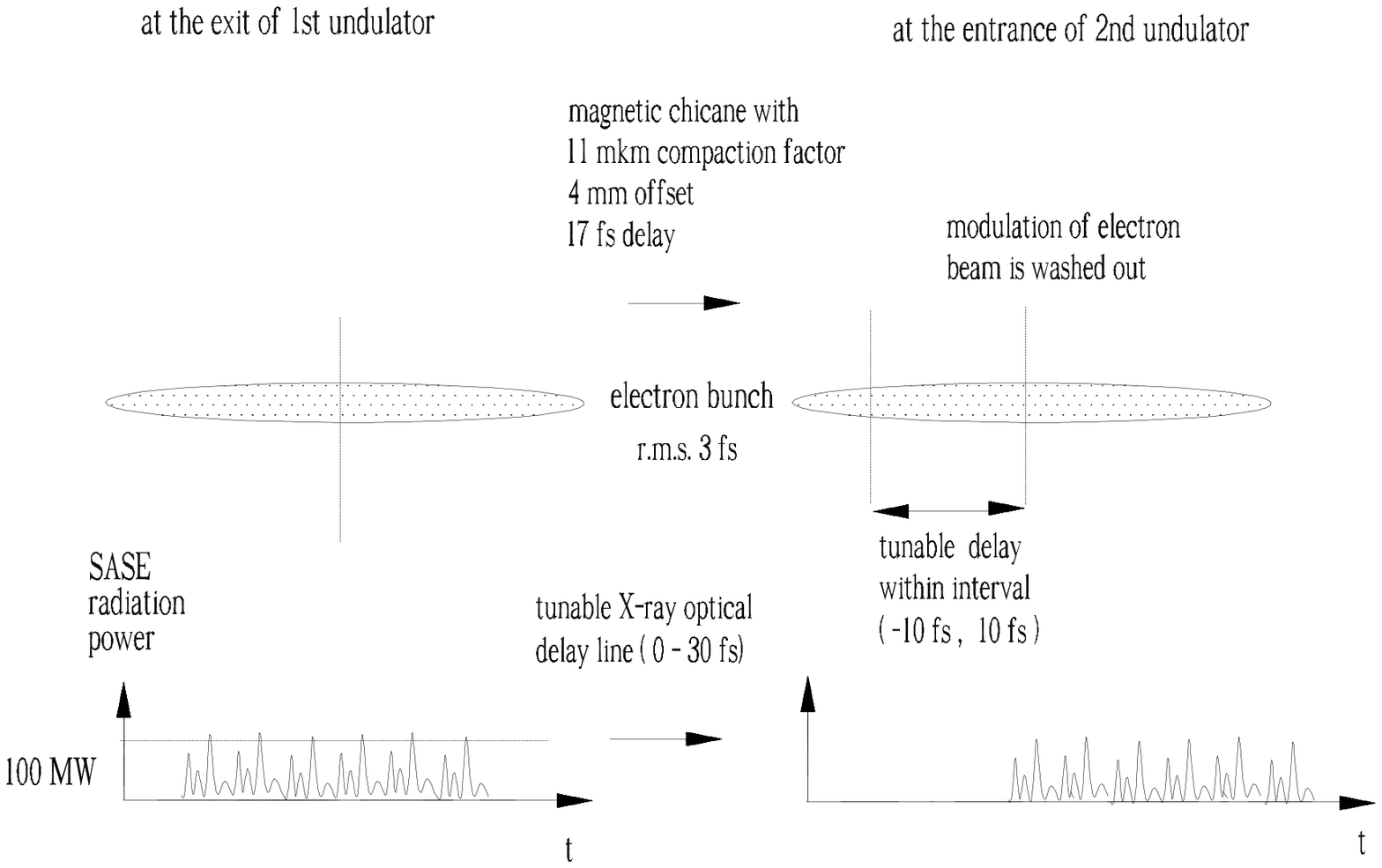}
\caption{Sketch of X-ray intensity autocorrelation measurement
using tunable X-ray optical delay line within magnetic chicane.
This autocorrelator produces complete autocorrelation trace.}
\label{mm6}
\end{figure}
After the chicane, the electron beam and the X-ray radiation pulse
produced in the first part of the undulator enter the second part
of undulator, which is resonant at the same wavelength. In the
second part of the undulator the first X-ray pulse acts as a seed
and overlaps to the lasing part of electron bunch. Therefore, the
output power rapidly grows up to the GW-level. First and second
undulator parts are identical and operate in the linear FEL
amplification regime. The relative delay between electron bunch
and seed X-ray pulse can be varied by the X-ray optical delay line
installed within the magnetic chicane, as illustrated in Fig.
\ref{mm6}. Within the 1D FEL theory, which is not too far from
reality for the SASE X-ray case with a large diffraction
parameter, one can write the shot-to-shot averaged power in the
pulse from the first part of the undulator as

\begin{eqnarray}
\left<P(t)\right> = P_0 \exp\left[2 L_w \mathrm{Re}
(\Lambda(t))\right] \label{P1}
\end{eqnarray}
where $P_0$ is the equivalent shot-nose power, $L_w=36$ m (6
cells) is the length of the two identical undulator parts and
$\mathrm{Re}(\Lambda(t))$ is the time-dependent\footnote{Here we
deal with a parametric amplifier, where the properties of the
active medium, i.e. the electron beam, depend on time.} field
growth-rate. Similarly, the shot-to-shot averaged power in the
pulse from the second part of the undulator, which is seeded with
$\left<P(t-\tau)\right>$, $\tau$ being the variable delay, can be
written as

\begin{eqnarray}
\left<P_2(t,\tau)\right> = \left<P(t-\tau)\right> \exp\left[2 L_w
\mathrm{Re} (\Lambda(t))\right] =
\frac{1}{P_0}\left<P(t-\tau)\right>\left<P(t)\right>~.\label{P2}
\end{eqnarray}
The subsequent measurement procedure consists in recording the
shot-to-shot averaged energy per pulse at the exit of the second
part of the undulator as a function of the relative delay between
electron bunch and seed X-ray pulse, with the help of an
integrating photodetector. This yields the autocorrelation
function

\begin{eqnarray}
A(\tau) = \int_{-\infty}^{\infty} dt
\left<P(t-\tau)\right>\left<P(t)\right>~.\label{AC}
\end{eqnarray}
Autocorrelation measurements are well known methods in laser
physics. Early on, it was realized that the only event fast enough
to measure an ultrashort pulse is the pulse itself.  A number of
schemes have been developed over the past decades to better
measure ultrashort laser pulses. Most of them have been
experimental implementations and variations of autocorrelators,
i.e. devices capable of measuring the autocorrelation function of
a given pulse, Eq. (\ref{AC}). Our scheme actually provides a
device capable of performing an intensity autocorrelation
measurement.

\begin{figure}[tb]
\includegraphics[width=1.0\textwidth]{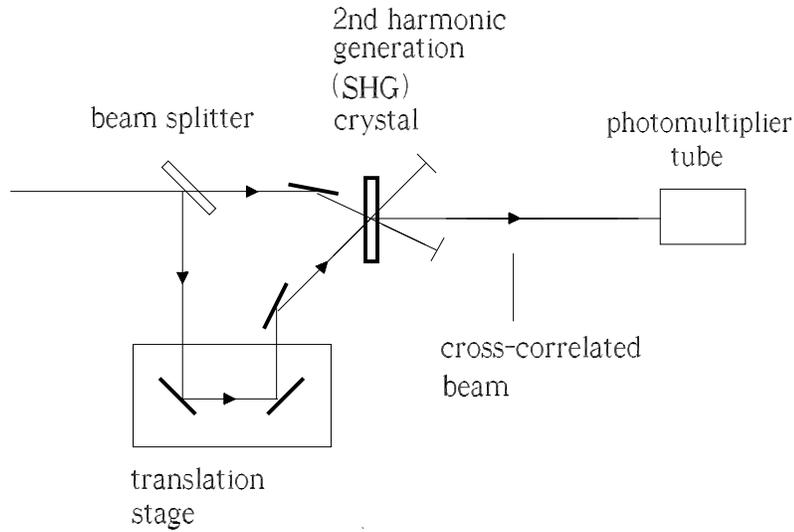}
\caption{Experimental layout for measuring the optical pulse
intensity versus time. At optical frequencies, an intensity
autocorrelator using second-harmonic generation can be used. A
pulse is split into two, one is variably delayed with respect to
other, and the two pulses are overlapped in a second harmonic
crystal. The second harmonic pulse energy is measured versus the
delay, giving the autocorrelation trace.} \label{mm2}
\end{figure}
Note that, in order to perform  intensity autocorrelation
measurements, one should insert a nonlinear element into an
interferometer. In ultrashort laser physics, the most common
approach in the visible range involves second-harmonic generation
(SHG), in which a nonlinear crystal is used to generate light at
twice the input optical frequency (see Fig. \ref{mm2}). The
measurement procedure is to record the time-averaged second
harmonic pulse energy as a function of the relative delay $\tau$
between the two identical versions of the input pulse. Due to
nonlinearity, the total energy in the second-harmonic pulse is
greater when the two pulses incident on the nonlinear crystal
overlap in time. Therefore, the peak in the second-harmonic power
plotted as a function of $\tau$ contains information about the
pulse width.

Similarly, here we assumed that we dispose of an ensemble of
identical pulses, so that the autocorrelation function can be
constructed from a large number of energy measurements taken for a
different delay parameter $\tau$. The measured energy is the sum
of a constant background term due to startup from shot noise from
each part of the undulator, and of the intensity autocorrelation
term, which arises from the interaction of the delayed electron
bunch and the seed SASE pulse from the first part of undulator.
Due to high-gain FEL amplification in the second part of the
undulator, the total energy in the X-ray pulse at the exit of the
setup is much higher when the seed SASE pulse and the electron
bunch overlap in time. This means that we effectively deal with a
background-free intensity autocorrelation function measurement.
Therefore, the peak in the shot-to-shot averaged energy of the
X-ray pulse at the setup exit, plotted as a function of $\tau$,
contains information about the averaged X-ray pulse width.

One immediately recognizes the physical meaning of the
autocorrelation function. The Fourier transform of the
autocorrelation function, $\bar{A}(\omega)$, is related to the
Fourier transform of the signal function $W(\omega)$, i.e. to the
Fourier Transform of the intensity vs time, by $\bar{A}(\omega) =
|W(\omega)|^2$. An autocorrelation function is always a symmetric
function. Thus, $\bar{A}(\omega)$ is a real function, consistent
with a symmetric function in the time domain. The intensity
autocorrelation function assumes its maximum value at $\tau = 0$.
Moreover, the autocorrelation function is an even function of
$\tau$, independently of the symmetry of the actual pulse.
Therefore, one cannot uniquely recover the pulse intensity profile
from the knowledge of the autocorrelation function only. This is
also understandable from the fact that correlation techniques
provide the possibility to measure the modulus of the Fourier
transform of the signal function, while information about its
phase is missing.

However, since the pulses exhibit no overlap for delays much
longer than the pulse width, the autocorrelation function goes to
zero for values of $\tau$ larger than the pulse width. Therefore,
the width of the correlation peak gives information about the
pulse width. One can estimate the FWHM of the radiation pulse from
the knowledge of the autocorrelation function if one assumes a
specific pulse shape. Then, the FWHM can be found by dividing the
intensity autocorrelation FWHM by a deconvolution factor, which is
specific for a given shape. If one deals with smoothly varying
pulse shapes, the deconvolution factor is about $1.5$
\cite{DIEL,WIEN} and the variation in the deconvolution factor is
of the order of $10 \%$ only. Therefore, the pulse duration can be
approximatively obtained from the knowledge of the FWHM of the
intensity autocorrelation function, even though the pulse shape
remains unknown.

\begin{table}
\caption{Parameters for the short pulse mode used in this paper.}

\begin{small}\begin{tabular}{ l c c}
\hline
& ~ Units &  Short pulse mode \\
\hline
Undulator period      & mm                  & 35.6   \\
I and II stage length & m                   & 35.9   \\
Segment length        & m                   & 6.00    \\
Segments per stage    & -                   & 6     \\
K parameter (rms)     & -                   & 2.9805  \\
$\beta$               & m                   & 27     \\
Wavelength            & nm                  & 0.15   \\
Energy                & GeV                 & 17.5   \\
Charge                & nC                  & 0.025  \\
Bunch length (rms)    & $\mu$m              & 1.0    \\
Normalized emittance  & mm~mrad             & 0.4    \\
Energy spread         & MeV                 & 1.5    \\
\hline
\end{tabular}\end{small}
\label{tab:fel-par}
\end{table}
In the next Section we will present a feasibility study of our
method, based on simulations with the code Genesis 1.3
\cite{GENE}, with the help of parameters in Table
\ref{tab:fel-par}. We further discuss an alternative method for
radiation pulse width measurement which is based on simpler
hardware, but should rely on trace retrieval algorithms to compute
the full autocorrelation trace from a-priori knowledge of the
electron bunch properties.

\section{Feasibility study}

\begin{figure}[tb]
\includegraphics[width=1.0\textwidth]{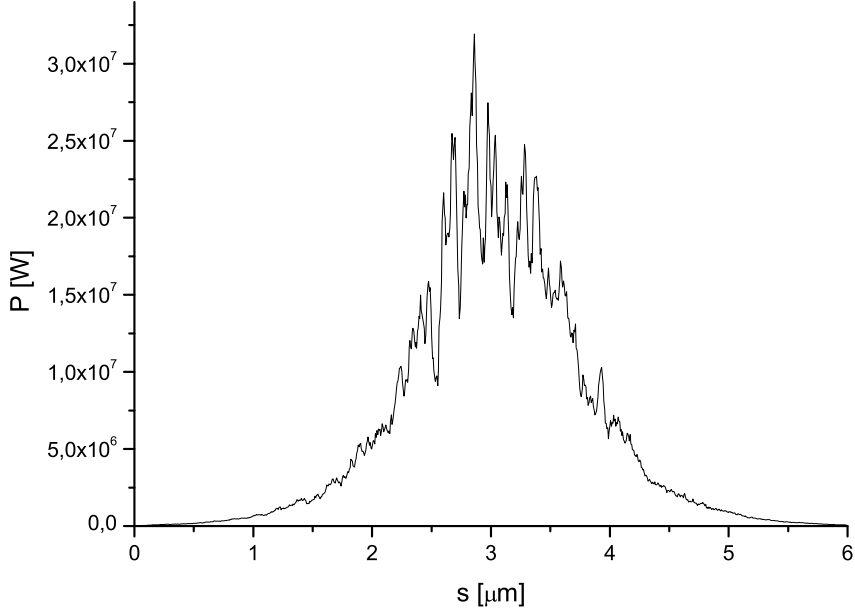}
\caption{SASE beam power distribution (single shot) after the
first undulator part ($36$ m-long).} \label{mm1b}
\end{figure}

First we let the electron beam through the first part of the
undulator, which is $36$ m long and is resonant at $0.15$ nm. A
picture of a single-shot beam power distribution after the first
part of the undulator is shown\footnote{Note that the Genesis
output consist in the total power integrated over full grid up to
an artificial boundary without any spectral selection. Therefore,
Fig. \ref{mm1b} includes a relatively large spontaneous emission
background, which has a much larger spectral width. The second
stage of our setup automatically selects the coherent contribution
within narrow bandwidth only. As a result, the simulation of the
average power profile at the exit of the first stage can only be
performed with the help of a special post processing filter.} in
Fig. \ref{mm1b}. If one would make an average over many shots, one
would obtain $\left<P(t)\right>$ given in Eq. (\ref{P1}).

\begin{figure}[tb]
\includegraphics[width=1.0\textwidth]{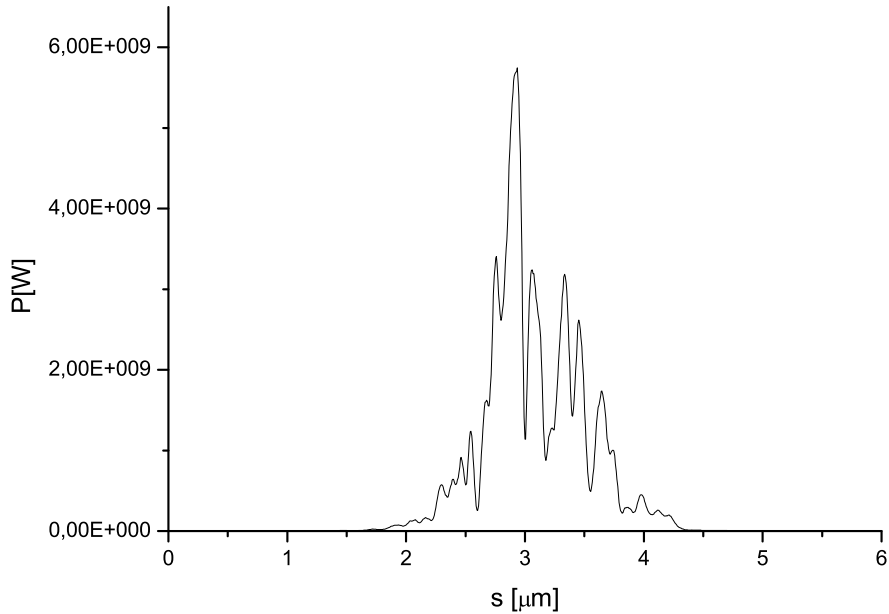}
\caption{Beam power distribution (single shot) after the second
part of the undulator ($36$ m-long)single shot. The SASE seeded
power distribution in this plot is obtained with a fresh beam and
for zero delay.} \label{mm2b}
\end{figure}

%
After the first part of the undulator, the electron beam is
delayed relatively to the photon beam, of a continuously tunable
temporal interval $\tau$, as illustrated in Fig. \ref{mm6}.
Moreover, the microbunching produced in the first part of the
undulator is washed out. Energy spread and energy loss induced
during the linear process in the first part of the undulator are
taken into account, but they are small, and the electron beam can
still undergo the SASE process in the second part of the
undulator. At this point, the radiation pulse $P_2(t,\tau)$ is
produced. A picture of a single-shot beam power distribution after
the second part of the undulator is shown in Fig. \ref{mm2b}. If
one would make an average over many shots of this figure, one
would obtain $\left<P_2(t,\tau)\right>$ given in Eq. (\ref{P2}).
We performed averaging over $10$ shots for each value of $\tau$.

\begin{figure}[tb]
\includegraphics[width=1.0\textwidth]{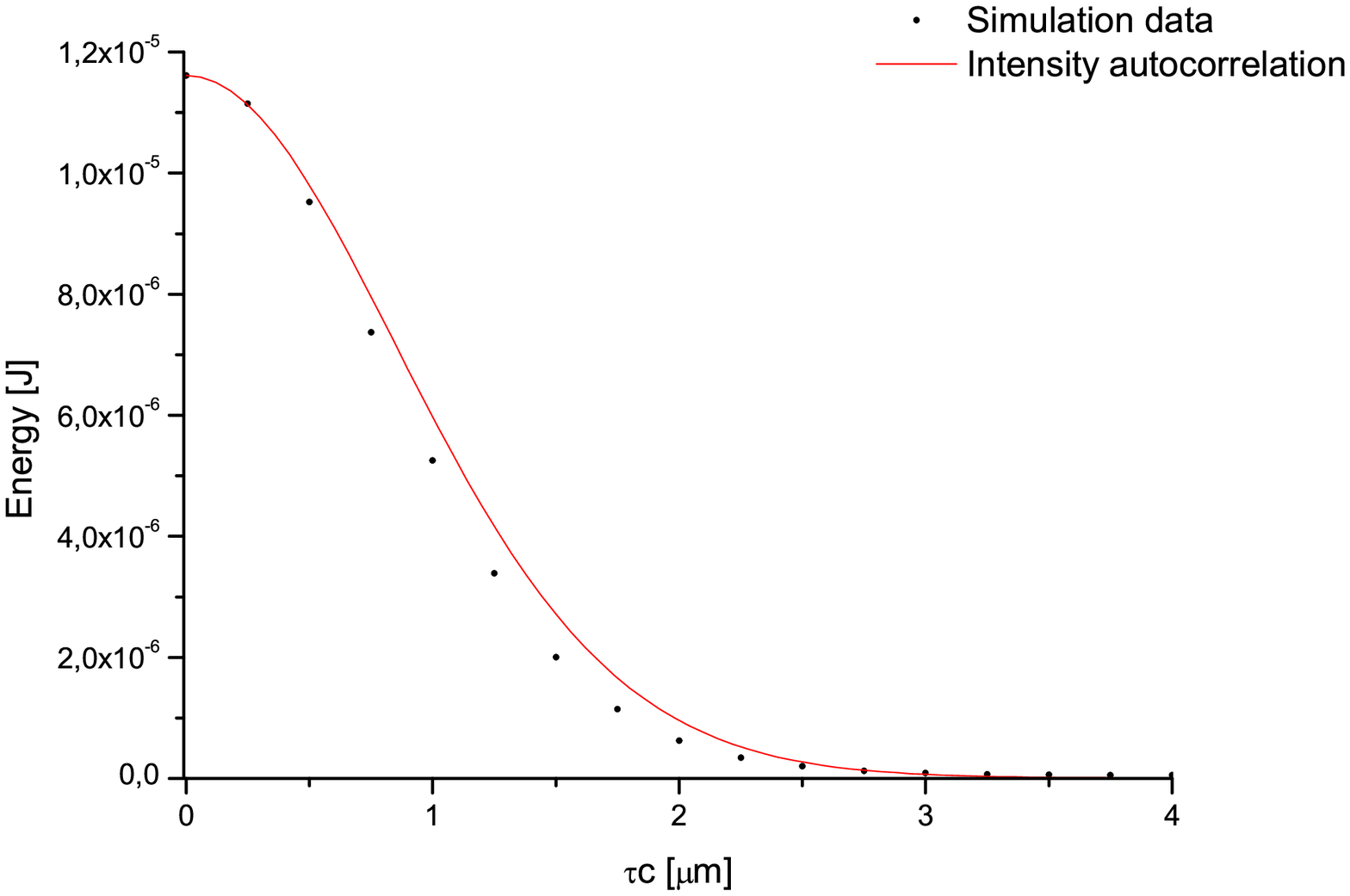}
\caption{Energy per pulse recorded at the integrating detector as
a function of the delay $\tau$. It constitutes the intensity
autocorrelation trace.} \label{mm3b}
\end{figure}

As discussed before, in order to obtain the intensity
autocorrelation function, after having calculated
$\left<P_2(t,\tau)\right>$ for different values of $\tau$ we need
to integrate it in time. The result is the average energy in the
pulse at the exit of the second undulator as a function of $\tau$
calculated with a 3D FEL code, i.e. the autocorrelation trace,
which is shown in Fig. \ref{mm3b} (black circles). Averaging have
been performed over ten shots.

\begin{figure}[tb]
\includegraphics[width=1.0\textwidth]{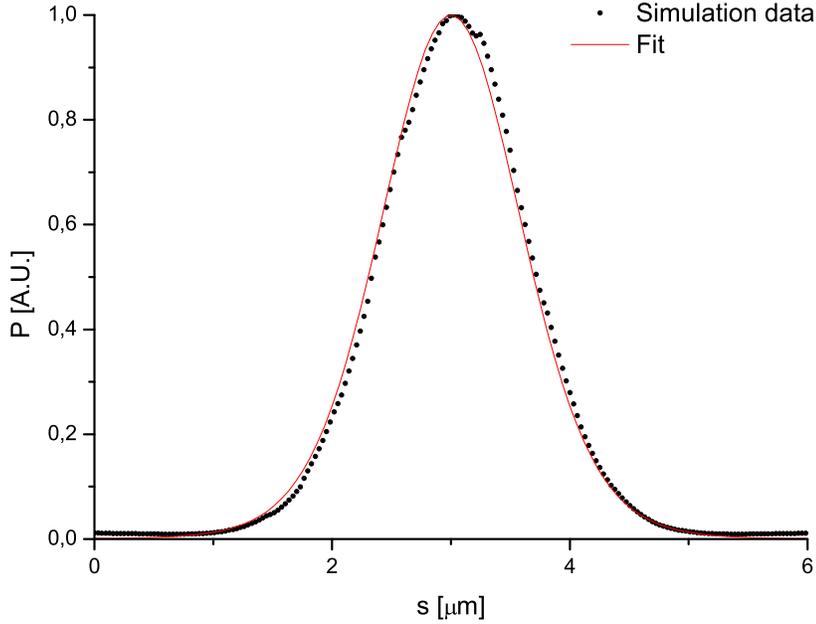}
\caption{Normalized gain envelope after the first undulator in the
case of startup from constant laser power. This plot is identical
to the envelope of the average power in the case of startup from
shot noise.} \label{mm4b}
\end{figure}

As shown before, in the 1D approximation, the autocorrelation
trace calculated in this way should just coincide with the
autocorrelation of the average power (i.e. the gain profile) for
the first undulator stage. We can independently calculate such
gain profile. Namely, instead of considering the start-up from
noise in the first undulator, we can simulate the case when a
constant laser power is fed at the entrance of the FEL and no shot
noise is considered. As a result, we obtain $\left<P(t)\right>$,
the ensemble average of the power distribution after the first
undulator for startup from shot noise, i.e. the gain envelope.
This result is plotted in Fig. \ref{mm4b} (black circles). The
expected functional dependence of $\left<P\right>$ on time is
given in Eq. (\ref{P1}) and, once normalized to unity, it can be
equivalently written as

\begin{eqnarray}
\left<P(t)\right> = \exp\left[\alpha I^{1/3}(t)\right]~,
\label{PI}
\end{eqnarray}
where $I(t)$ is the current, also normalized to unity for
simplicity, $I(t)=\exp[-(t-t_0)^2/(2 \sigma^2)]$. Since $\sigma$
and $t_0$ are known, Eq. (\ref{PI}) can be used to fit the
simulation data, with $\alpha$ as the only free parameter. It
turns out that the best fit, shown with a solid line in Fig.
\ref{mm4b}, is for $\alpha = 8.9$.

Since the best fit value for $\alpha$ is now fixed, we can
calculate the autocorrelation trace using Eq. (\ref{AC}). The
result is plotted with a solid line in Fig. \ref{mm3b}. It is seen
that there is a good agreement between actual intensity
autocorrelation and the black circles.  Deviations can be due to
differences between the 1D and the 3D treatments, to the fact that
in the second stage we work near to the non-linear regime, that we
neglect the temporal dependence of $P_0$, or to the fact that for
these exemplifications we used only $10$ shot averaging. However,
this accuracy is sufficient for our purpose of demonstrating the
feasibility of the method.

Now, let us suppose that we do not know the gain curve, but we
simply measure the energy per pulse from our setup, i.e. the black
circles in Fig. \ref{mm3b}. By inspecting this autocorrelation
trace, we conclude that the FWHM of the autocorrelation function
is about $1.9 \mu$m. Assuming, as discussed above, a deconvolution
factor of $1.5$, we obtain an estimate for the FWHM of the
radiation pulse of about $1.3~\mu$m. The actual FWHM of the
average power distribution from Fig. \ref{mm4b} is, instead, of
$1.4~\mu$m. This gives an idea of the accuracy of the estimation
of the radiation profile width with the deconvolution factor
$1.5$.

\section{Simplest measurement using a magnetic chicane only}

\begin{figure}[tb]
\includegraphics[width=1.0\textwidth]{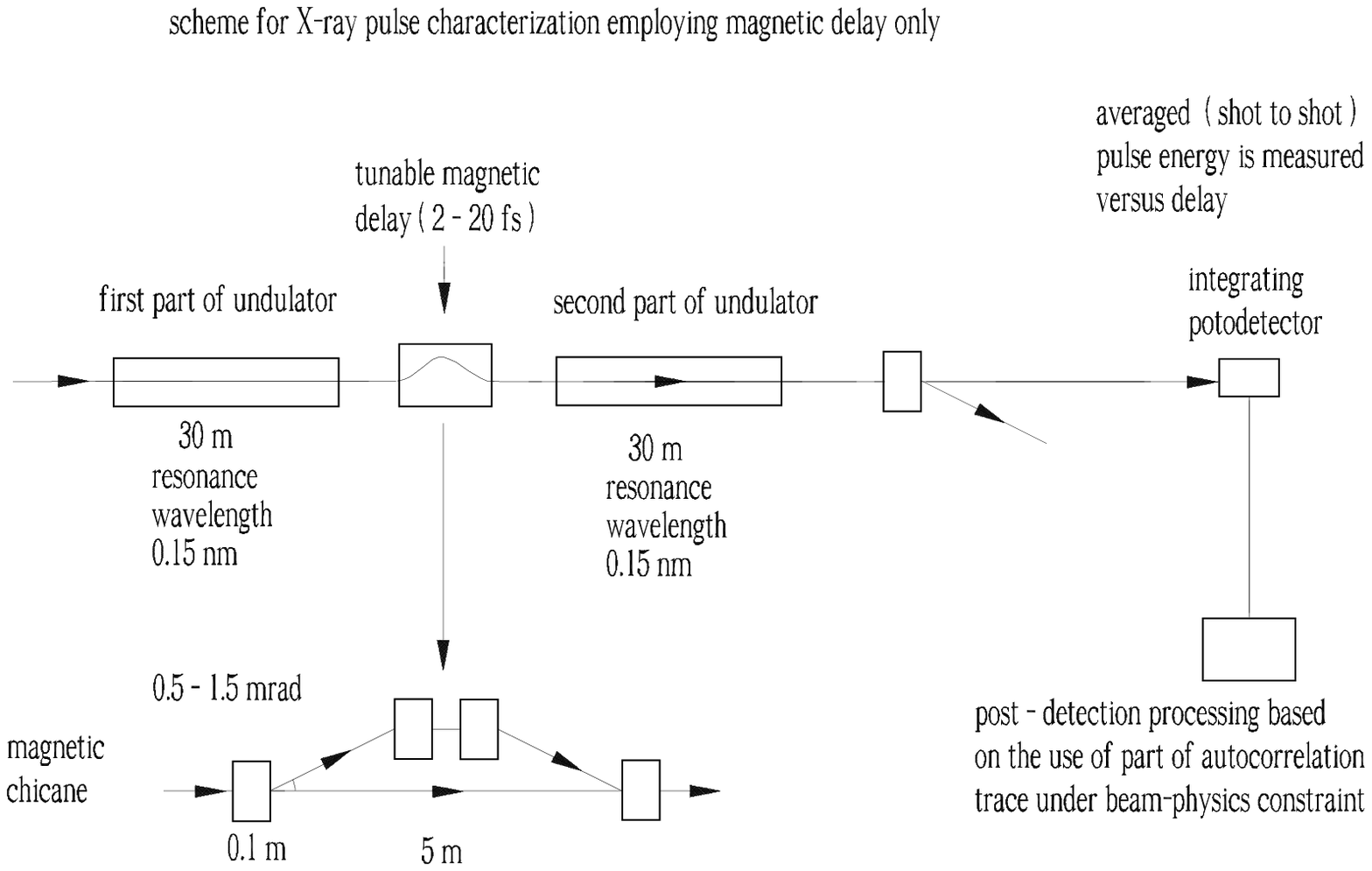}
\caption{Simplest ultra-short X-ray pulse-measurement setup which
utilizes magnetic-chicane delay line only.  The setup cannot
provide full autocorrelation trace. The compaction factor of
magnetic chicane must be large enough ( $> 1 \mu$m ) for
performing "fresh" bunch technique. The  trace retrieval is based
on computer simulations  and a-priory information about electron
bunch properties.} \label{mm3}
\end{figure}

\begin{figure}[tb]
\includegraphics[width=1.0\textwidth]{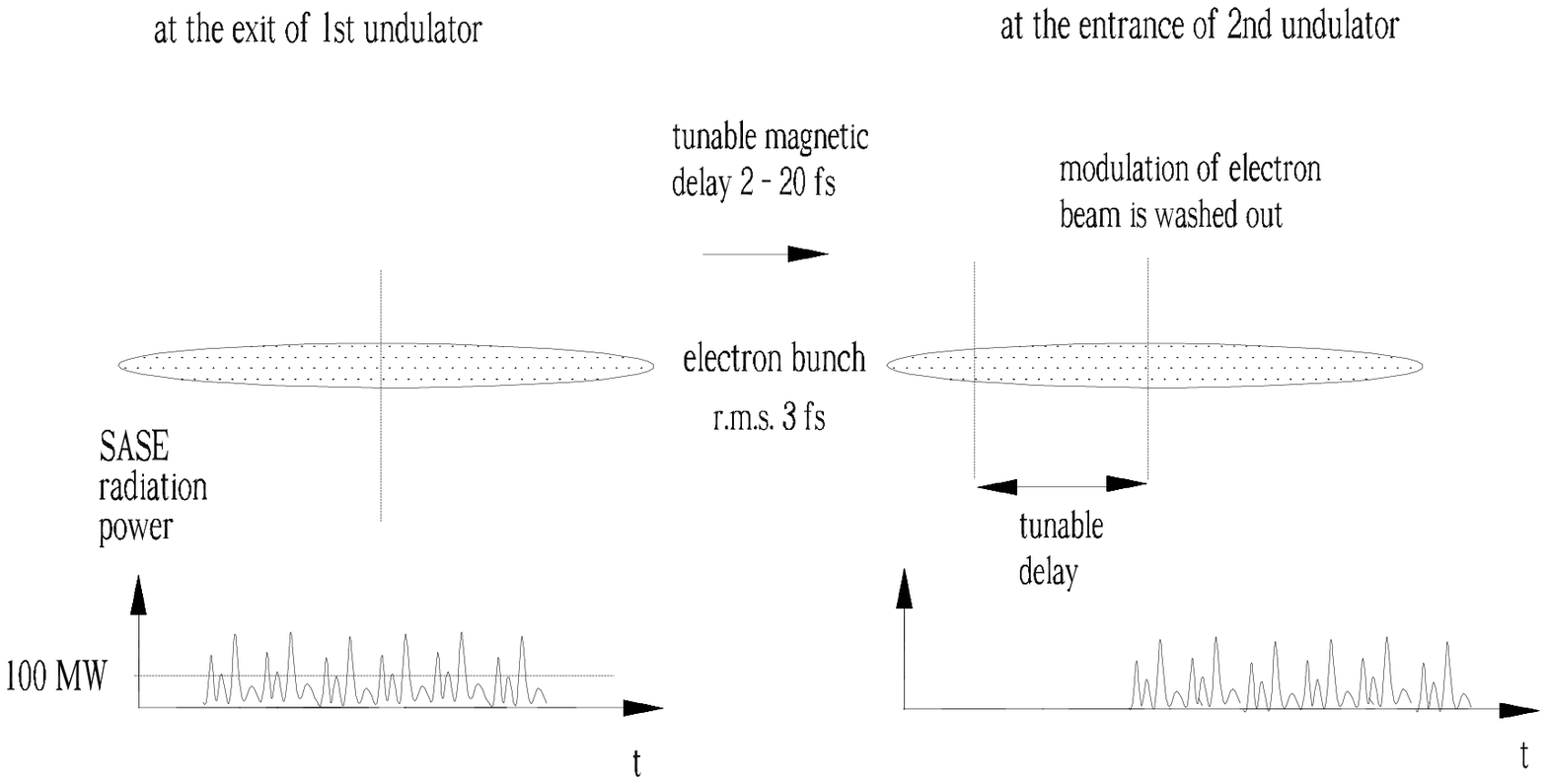}
\caption{Sketch of X-ray intensity autocorrelation measurement
using tunable magnetic delay. This autocorrelator produces part of
autocorrelation trace only.} \label{mm5}
\end{figure}

As an alternative to the method considered above, we also propose
a technique to measure the width of ultrafast radiation pulses
based on simpler hardware. The idea is to rely on a magnetic
chicane only, without an optical delay line, and is illustrated in
Fig. \ref{mm3}. Also a magnetic chicane alone, in fact, can
provide a delay of the electron beam relative to the radiation
pulse, as illustrated in Fig. \ref{mm5}. However, the compaction
factor of the magnetic chicane must always obey to the constraint
to be large enough to allow for the microbunching produced in the
first part of the undulator to be washed out. Therefore, the delay
$\tau$ cannot be set to zero as in the previous case, and the
presence of a simpler hardware is paid by the fact that the setup
cannot provide a full autocorrelation trace.

It is therefore necessary to recover the missing data with the
help of computer simulations and information about the electron
bunch properties, which is available from other measurements.

\section{Conclusions}

In this paper we presented a novel method to measure the width of
ultrashort femtosecond pulses of x-rays produced in XFELs. The
technique is based on the measurement of the average intensity
autocorrelation function of the x-ray pulse. To this purpose, we
use a fresh-bunch technique. Measurement of the energy in the
radiation pulse produced in our setup yields the autocorrelation
function. The method works with limited hardware, in its simplest
form with a weak magnetic chicane installed in place of one of the
undulator segments. The magnetic chicane can be installed without
any perturbation of the XFEL focusing structure, and does not
disturb the baseline undulator mode. Our proposal is therefore
cheap, robust, and does not present any risk for the functionality
of the facility.

\section{Acknowledgements}

We are grateful to Massimo Altarelli, Reinhard Brinkmann, Serguei
Molodtsov and Edgar Weckert for their support and their interest
during the compilation of this work.

\end{document}